\documentclass[onecolumn,12pt]{IEEEtran}
\usepackage{amsmath, graphics,amssymb,epsfig,subfigure}

\usepackage{amsbsy}
\usepackage{graphicx}
\usepackage{graphics}
\usepackage{algorithm}
\usepackage{algpseudocode}
\usepackage{subfigure}
\usepackage{cite}
\usepackage{slashbox}
\usepackage{multirow}
\usepackage{ctable}
\usepackage{fancyhdr}
\usepackage{lastpage}
\usepackage{setspace}
\usepackage[pagewise]{lineno}

\newcommand{\cB}{{\mathcal B}}

\newcommand{\cF}{{\mathcal F}}

\newcommand{\cI}{{\mathcal I}}

\newcommand{\cM}{{\mathcal M}}

\newcommand{\cS}{{\mathcal S}}

\newcommand{\Hv}{\mbox{$\bf H $}}

\newcommand{\xv}{\mbox{$\bf x $}}

\newcommand{\Pv}{\mbox{$\bf P $}}

\newcommand{\nv}{\mbox{$\bf n $}}

\newcommand{\hv}{\mbox{$\bf h $}}

\newcommand{\Iv}{\mbox{$\bf I $}}

\newcommand{\Gv}{\mbox{$\bf G $}}

\newcommand{\rv}{\mbox{$\bf r $}}

\newcommand{\be}{\begin{equation}}
\newcommand{\ee}{\end{equation}}
\newcommand{\bea}{\begin{eqnarray}}
\newcommand{\eea}{\end{eqnarray}}
\newcommand{\bdp}{\begin{displaymath}}
\newcommand{\edp}{\end{displaymath}}

\begin{document}
\title{Link Performance Abstraction for Interference-Aware Communications (IAC)}
\author{\authorblockN{Heunchul Lee, Taeyoon Kim,  Wonwoo Park, and Jonghan Lim } \\
\authorblockA{The Corporate R$\text{\&}$D Center, Samsung Electronics\\
 email: \{heunchul.lee, taeyoon1.kim,  wonwoo1.park, jonghan.lim\}@samsung.com} 
}
\maketitle \thispagestyle{empty}
\begin{abstract}
Advanced co-channel interference aware signal detection has drawn research attention during the recent development of Long Term Evolution-Advanced (LTE-A) systems \cite{Bea:12}\cite{Jungwon:12} and the interference-aware communications (IAC) is currently being studied by 3GPP \cite{MediaTek:13}. This paper investigates link performance abstraction for the IAC systems employing maximum-likelihood detector (MLD). The link performance of MLD can be estimated by combining two performance bounds, namely, linear receiver and genie-aided maximum-likelihood (ML) receiver.  It is shown that the conventional static approach based on static parameterization \cite{Moon:12}, while working well under  moderate and weak interference  conditions, fails to generate a well-behaved solution in the strong interference case. Inspired by this observation, we propose a new adaptive approach where the combining parameter is adaptively adjusted according to instantaneous interference-to-signal ratio (ISR). The basic idea is to exploit the probabilistic behavior of the optimal combining ratio over the ISR.
The link-level simulation results are provided to verify the prediction accuracy of the proposed link abstraction method. Moreover, we use the proposed link abstraction model as a link-to-system interface mapping in system-level simulations to demonstrate the performance of the IAC receiver in interference-limited LTE systems.\end{abstract}
\section{Introduction}
In order to improve coverage and spectral efficiency,  Long Term Evolution-Advanced (LTE-A) systems are designed to operate with an aggressive frequency reuse factor and  a high density of base station (BS) sites. Especially, the BSs in the LTE-A network may achieve multiple accesses via  multiple-input multiple-output (MIMO) technology and allow user equipments (UEs) to share the same frequency-time resources. This will inevitably lead to severe inter-cell interference problem. In this context, advanced features such as coordinated multipoint (CoMP) transmission and  enhanced intercell interference coordination (eICIC) have been specified in recent LTE releases to improve cell-edge throughput for the interference-limited scenarios \cite{3GPP:11} \cite{dahlman20114g}.
More recently, advanced co-channel interference aware signal detection has drawn research attention during the recent development of LTE-A systems \cite{Bea:12}\cite{Jungwon:12}. The interference-aware communications (IAC), termed network-assisted interference  cancellation and suppression (NAICS), is currently being studied for inclusion in LTE Release 12 \cite{MediaTek:13}. 

With network assistance, the advanced IAC receiver will provide significant performance benefits  \cite{Jungwon:12}\cite{MediaTek:13}. In order to realize the actual performance improvements in LTE-A systems, it is essential to incorporate the performance gain from employing the IAC receiver into adaptive transmission techniques such as link adaptation. Channel quality indicator (CQI) plays a key role in the link adaptation process. The link adaptation is performed by the BS using CQI reports from a UE. This means that more accurate CQI measurement at the UE side gives more throughput gain. To this end, the UE requires a link abstraction method to estimate the block error rate (BLER) of each modulation and coding scheme (MCS) for given current channel conditions and selects as a CQI value the highest MCS which achieves a target  BLER. 
In addition, the link abstraction methodology provides a physical-layer (PHY) abstraction as a link-to-system interface mapping in system-level simulations (SLSs). Note that evaluating the system-level performance of different air-interface technologies requires the use of instantaneous BLER for a given channel realization rather than long-term BLER in a fading channel. In summary, link abstraction methods should be able to accurately predict time-varying BLER of a given link without extensive simulation.

The key technologies in LTE systems are orthogonal frequency division multiplexing (OFDM) and MIMO. The OFDM modulation technique divides the total available bandwidth into a number of equally spaced subcarriers, 
resulting in different fading gains for different subcarriers. Furthermore, bit-interleaved coded modulation (BICM) is considered for increasing the code diversity on fading channels \cite{Zehavi:92}\cite{Caire:98}\cite{Inkyu:06}. In this paper, we investigate link performance abstraction of the IAC employing the maximum-likelihood detection (MLD)  for MIMO-OFDM systems in multicell multiuser interfering networks.  

Traditionally, signal-to-interference-plus-noise ratio (SINR) is used as a representative output to obtain an instantaneous BLER \cite{Roshni:08}. Taking account of the fact that the coded bits transmitted by MIMO-OFDM systems are spread over different spatial layers and subcarriers, a link abstraction method of MIMO-OFDM systems can be composed of two stages, namely, the layer separation in the MIMO system and effective SINR mapping (ESM) in the OFDM block. First, at each OFDM subcarrier, we derive a post-processing SINR for each spatial layer of a MIMO system and then utilize mutual information per coded bit (MIB) metric to convert a set of different post-processing SINRs, obtained over the frequency-selective coded OFDM system, into a single MIB. This MIB value is used to predict instantaneous BLER of MIMO-OFDM systems. Meanwhile, we use as the reference curves the BLER curves generated for all MCSs under additive white Gaussian noise (AWGN) assumption. 

The post-processing SINR of each spatial layer is dependent on the detection algorithm used in MIMO systems.  
In the case of MIMO systems with linear receivers such as  minimum mean-squared error (MMSE) and zero-forcing (ZF) receivers, the post-processing SINR is readily given by the output SINR.
However, when it comes to the  maximum-likelihood (ML) receiver, it is not straightforward to calculate the post-processing SINR since the ML-based demodulation is a non-linear process.
In \cite{Moon:12}, a new approach was introduced to estimate the post-MLD SINR for a single cell MIMO by combining the two performance bounds of linear MMSE receiver and genie-aided ML receiver. 
Unfortunately, the previous approach assumes the combining ratio between the two bounds to remain fixed for the involved MCS. It is shown in this paper that the conventional static combining approach, while working well under moderate and weak interference conditions, fails to generate a well-behaved solution in the strong interference case. Inspired by this observation, we propose an adaptive approach where the combining parameter is adaptively adjusted based on the instantaneous interference-to-signal ratio\footnote{In this paper, we use the term interference-to-signal ratio (ISR) instead of signal-to-interference ratio (SIR) to emphasize that as will become clear later in this paper, the achievable MIB of MLD-based IAC receiver increases proportionally to the ISR.} (ISR). The basic idea is to exploit the probabilistic behavior of the optimal combining ratio over the ISR. The link-level simulation (LLS) results are provided to verify the prediction accuracy of the proposed link abstraction method. We are also interested in gaining insight in the potential gains of using the IAC receiver compared to the baseline LTE  receiver defined in \cite{3GPP:829}.
We use the proposed link abstraction model as a link-to-system interface mapping in system-level simulations to demonstrate the performance of the IAC receiver in the interference-limited LTE systems

The remainder of this paper is organized as follows: first, Section
\ref{sec:2} presents an overview of IAC concept and introduces the equations for the exact evaluation of MIB. A brief review of the conventional static approach to the layer separation is provided in Section \ref{sec:3}. Meanwhile, we show that the lower bound based on the linear MMSE receiver results in misleading lower bound in the conventional static approach under strong interference conditions.
In Section \ref{sec:4}, we propose an ISR-adaptive approach to overcome the drawback of lower bound. Section \ref{sec:5} shows simulation results comparing with the conventional approach. Finally, conclusions are made in Section \ref{sec:6}.
\section{IAC and Mutual Information} \label{sec:2}
In this section, we describe system model and the achievable MIB of IAC receiver. To this end, we consider  downlink MIMO-OFDM systems in multicell environments where two BSs equipped with $N_t$ transmit antennas are transmitting their own messages, respectively, to the desired UEs equipped with $N_r$ receive antennas. 

Let us denote the $V_i$-dimensional complex signal vector
transmitted from BS $i$ at the $k$-th subcarrier as $\xv^i_k=\left[x^{i,1}_k,\cdots,x^{i,V_i}_k\right]^T$, $i=1,2$, where $x^{i,v}_k$ denotes the $v$-th spatial layer at subcarrier $k$, $V_i$ indicates the number of layers, and $(\cdot)^T$ denotes the transpose of a vector. Symbol $x^{i,v}_k$ is chosen from $M^i_c$-ary constellation set $\mathbb{S}_i$, i.e.,  $x^{i,v}_k \in \mathbb{S}_i$. Without loss of generality, we assume that BS $1$ is the serving BS and BS $2$ is the interfering BS.  
The channel model from BS $i$ to the desired UE at subcarrier $k$ is represented by an $N_r$-by-$N_t$ channel matrix  $\Gv^i_k$, whose $(p,q)$ entry denotes the path gain from antenna $q$ of BS $i$ to antenna $p$ at the UE, modeled as independent complex Gaussian random variables with zero mean and unit variance, i.e., Rayleigh fading. The average transmit power of $x^{i,v}_k$ is assumed to be normalized to one, i.e., $E[|x^{i,v}_k |^2]=1$, where $E[\cdot]$ denotes the expectation operator and $|\cdot|$ represents the absolute value of a complex number. 

Let us define $\rv_k$ as the $N_r$-dimensional complex received signal vector by the desired UE at the subcarrier $k$. Then, $\rv_k$ can be written as  
 \bea \label{eq:1}
\rv_k = \Hv^1_k \xv^1_k +\Hv^2_k \xv^2_k + \nv_k, ~~~~~~ \text{for}~~ k =
1,2,\cdots,K, \eea
where $\Hv^i_k$ denotes an effective channel matrix comprising distance dependent pathloss, the actual channel matrix $\Gv^i_k$ with size $N_r$-by-$N_t$ and the precoding matrix $\Pv^i_k$ with size $N_t$-by-$V_i$, $\nv_k$ denotes the additive noise
vector whose elements are independent and identically-distributed (i.i.d.)
complex Gaussian with variance $\sigma_n^2$ and $K$ represents the total number of coded subcarriers. 
Note that the actual transmitted signal vector is given  by $\Pv^i_k\xv^i_k$. Since the precoding matrix $\Pv^i_k$ of the LTE codebooks \cite{211:12} is normalized by the number of transmission layer $V_i$, we can define the average (per-user) signal-to-noise ratio (SNR) as $\frac{1}{\sigma^2_n}$.
 
Under our assumptions, the channel transition probability is given by
\bea  \label{eq:2}
p(\rv_k|\xv^1_k,\xv^2_k)
=\frac{1}{{(\pi\sigma^2_n)}^{N_r}} \exp{\left( -\frac{||\rv_k -
{\Hv}^1_k \xv^1_k - {\Hv}^2_k \xv^2_k ||^2}{\sigma^2_n}\right)}.
\eea

Let $b^{1}_{k,v,m}$ be the $m$th bit ($m=1,2,\cdots,\text{log}_2M^1_c$) of the constellation symbol $x^{1,v}_k$.
We denote $L(b^{1}_{k,v,m})$ as the log-likelihood ratio (LLR) value for bit $b^{1}_{k,v,m}$, which is
defined as 
\bea \label{eq:2b} L\left(b^{1}_{k,v,m}\right) =
\text{log}\frac{P\left(b^{1}_{k,v,m}=1\right)}{P\left(b^{1}_{k,v,m}=0\right)},
\eea 
where $P\left(b^{1}_{k,v,m}=b\right)$ denotes the probability
that the random variable $b^{1}_{k,v,m}$ takes on the value $b$,
$b=0$ or $1$.

The LLR in (\ref{eq:2b}) conditioned on the channel state information
can be rewritten as 
\bea \label{eq:4} L\left(b^{1}_{k,v,m}\right)
=\text{log}\frac{P\left(b^{1}_{k,v,m}=1|\rv_k,\Hv^1_k,\Hv^2_k\right)}{P\left(b^{1}_{k,v,m}=0
|\rv_k,\Hv^1_k,\Hv^2_k\right)}. \eea 

Then, assuming the interference-aware ML detection \cite{Jungwon:12} and i.i.d. uniform coded
bits $b^{1}_{k,v,m}$ without a priori information, we can compute the LLRs by 
\bea
\label{eq:5} L\left(b^{1}_{k,v,m}\right)
 =\text{log}\frac{\sum_{\xv^1_k
\in \chi_{1}^{v,m}(1)}\sum_{\xv^2_k \in \chi(2)}p(\rv_k|\xv^1_k,\xv^2_k)}
                {\sum_{\xv^1_k \in
\chi_{0}^{v,m}(1)}\sum_{\xv^2_k \in
\chi(2)}p(\rv_k|\xv^1_k,\xv^2_k)}, \eea 
where $\chi(i)$ denotes the set of all possible symbol vectors
$\xv^i_k$, which is obtained as the $V_i$-fold Cartesian product of
$\mathbb{S}_i$, and $\chi_b^{v,m}(1)$ denotes a set of all
symbol vectors $\in \chi(1)$ whose $b^{1}_{k,v,m}=b$, ($b=0$ or $1$).

BICM separates the MIMO detector and the decoder via a bit-level interleaver and each coded bit experiences a different quality of channel. Thanks to the interleaver, we assume that all the bits are independent. Then, by extending the results in \cite{Caire:98} and \cite{Biglieri:00a}, the mutual information of the bit channel for $b^{1}_{k,v,m}$ can be evaluated as  
\begin{align} \label{eq:3} 
&\cM^{MI}_{k,v,m}=1- \\ \nonumber
&\! E_{b,\xv^1_k,\xv^2_k,\rv_k,\Hv^1_k,\Hv^2_k}\!\left[\!\text{log}_2\frac{\sum_{\xv^1_k
\in \chi(1)}\sum_{\xv^2_k \in \chi(2)}p(\rv_k|\xv^1_k,\xv^2_k)}
                {\sum_{\xv^1_k \in
\chi_{d}^{v,m}(1)}\sum_{\xv^2_k \in
\chi(2)}p(\rv_k|\xv^1_k,\xv^2_k)}\!\right]\!. 
\end{align}

Finally, we arrive at the MIB of the $v$-th layer on the $k$-th subcarrier given by
\bea  \label{eq:4}
{\cM}^{ML}_{k,v}=\frac{\sum_{m=1}^{\log_2{M^1_c}}\cM^{MI}_{k,v,m}}{\log_2{M^1_c}}. 
\eea
The MIB is implicitly dependent on both the SNR and the interference-to-noise ratio (INR) from (\ref{eq:3}) and varies with the antenna configuration ($N_t$ and $N_r$) and the modulation level ($M^1_c$ and $M^2_c$).

The equations (\ref{eq:3}) and (\ref{eq:4}) will generate the exact MIB of the $v$-th layer on each subcarrier $k$. 
However, the main problem with this approach is that when evaluating
mutual information values in (\ref{eq:3}), the search candidate number of elements in $\chi(i)$  grows exponentially with the
number of transmit antennas and/or bits per symbol, which is prohibitively complex for practical use in link adaptation and SLS.
In the following sections, we consider a simple and computationally efficient approach for estimating the MIB of each spatial layer in MIMO systems.
\section{ Conventional static approach to the layer separation } \label{sec:3}
In this section, as the first stage of link performance abstraction of MIMO-OFDM systems, we present a brief review of the conventional static approach to the layer separation that derives a post-processing SINR of each spatial layer in MIMO systems. Meanwhile, we show that the lower bound based on the linear MMSE receiver results in misleading lower bound in the conventional static approach under strong interference conditions.

\subsection{Static Combining Approach}
As mentioned earlier, it is not straightforward to compute the post-processing SINR in the case of MIMO systems using MLD. We consider the layer separation method proposed in \cite{Moon:12}, where the post-MLD SINR is calculated as a function of the post-MMSE receiver SINR and the genie-aided interference-free (IF) receiver SINR, and extend it for the multicell MIMO downlink. 

As to unbiased MMSE receiver, the post-processing SINR of the $v$-th layer on the $k$-th subcarrier can be expressed as \cite{Stanford:379}
\bea  \label{eq:5}
\gamma^{MMSE}_{k,v}=\frac{1}{\sigma^2_{k,v}} -1, ~~~~~~ \text{for}~~ v =
1,2,\cdots,V_1,
\eea
where $\sigma^2_{k,v}$ denotes the mean-squared error (MSE) for the $v$-th layer at the $k$-th subcarrier. 

Under the assumptions made for the signal model in (\ref{eq:1}), the MSE, denoted as $\sigma^2_{k,v}$,  can be computed as 
\bea  \label{eq:6}
\sigma^2_{k,v}=\left[\left(\Iv_{N_t}+\frac{1}{\sigma^2_n}\bar{\Hv}_k^{\dagger}\bar{\Hv}_k \right)^{-1} \right]_{v,v},
\eea
where $\bar{\Hv}_k=\left[\Hv^1_k, \Hv^2_k\right]$, $(\cdot)^{\dagger}$ indicates the complex-conjugate transpose, $\Iv_r$ denotes an identity matrix of size $r$, and $[\cdot]_{n,n}$ represents the n-th diagonal element of a matrix.

In comparison, the post-MLD SINR can be upper-bounded by the genie-aided IF receiver and the corresponding SINR of the layer $v$ can be represented as
\bea  \label{eq:7}
\gamma^{IF}_{k,v} =\frac{||\hv^1_{k,v}||^2}{\sigma^2_n}, ~~~~~~ \text{for}~~ v =
1,2,\cdots,V_1,
\eea
where $\hv^i_{k,v}$ indicates the $v$-th column of $\Hv^i_k$.

By using the two SINRs given by (\ref{eq:5}) and (\ref{eq:7}), the post-MLD SINR can be lower-and-upper bounded as follows:
\bea  \label{eq:8}
\cF \left(\gamma^{MMSE}_{k,v}\right)\le \cF \left(\gamma^{ML}_{k,v}\right) \le \cF \left(\gamma^{IF}_{k,v}\right),
\eea
where the function $\cF$ can utilize different metrics such as channel capacity or MIB, i.e., $\cF(\gamma)=\log_2\left(\gamma+1\right)$ or  $\cF(\gamma)=\cI_{M^1_c}\left(\gamma\right)$. In this work, we focus on the MIB metric.

Here  $\cI_{M_c}\left(\gamma\right)$ denotes the MIB mapping function of SNR $\gamma$ for the involved modulation level $M_c$ under the assumption of AWGN channel. 
Unlike the metric of channel capacity, the MIB is the constellation-constrained capacity which is dependent on the signal constellation and the bit labeling. An efficient approach for MIB computation is developed in \cite{Roshni:08} by approximating the probability density function (PDF) of LLR with a mixture of Gaussian PDFs.   

In \cite{Moon:12}, it was proposed that the post-MLD SINR can be modeled by using a fixed combining ratio $\beta$ as
\bea  \label{eq:9}
\cF \left(\gamma^{ML}_{k,v}\right)=(1-\beta)\cF \left(\gamma^{MMSE}_{k,v}\right)+ \beta\cF \left(\gamma^{IF}_{k,v}\right),
\eea
where $\beta$ is a constant value for optimization.

\subsection{Behavior of $\gamma^{MMSE}_{k,v}$ and $\gamma^{IF}_{k,v}$ in interference environments} \label{sec:3B}

\begin{figure}
\begin{center}
\includegraphics[width=5.5in]{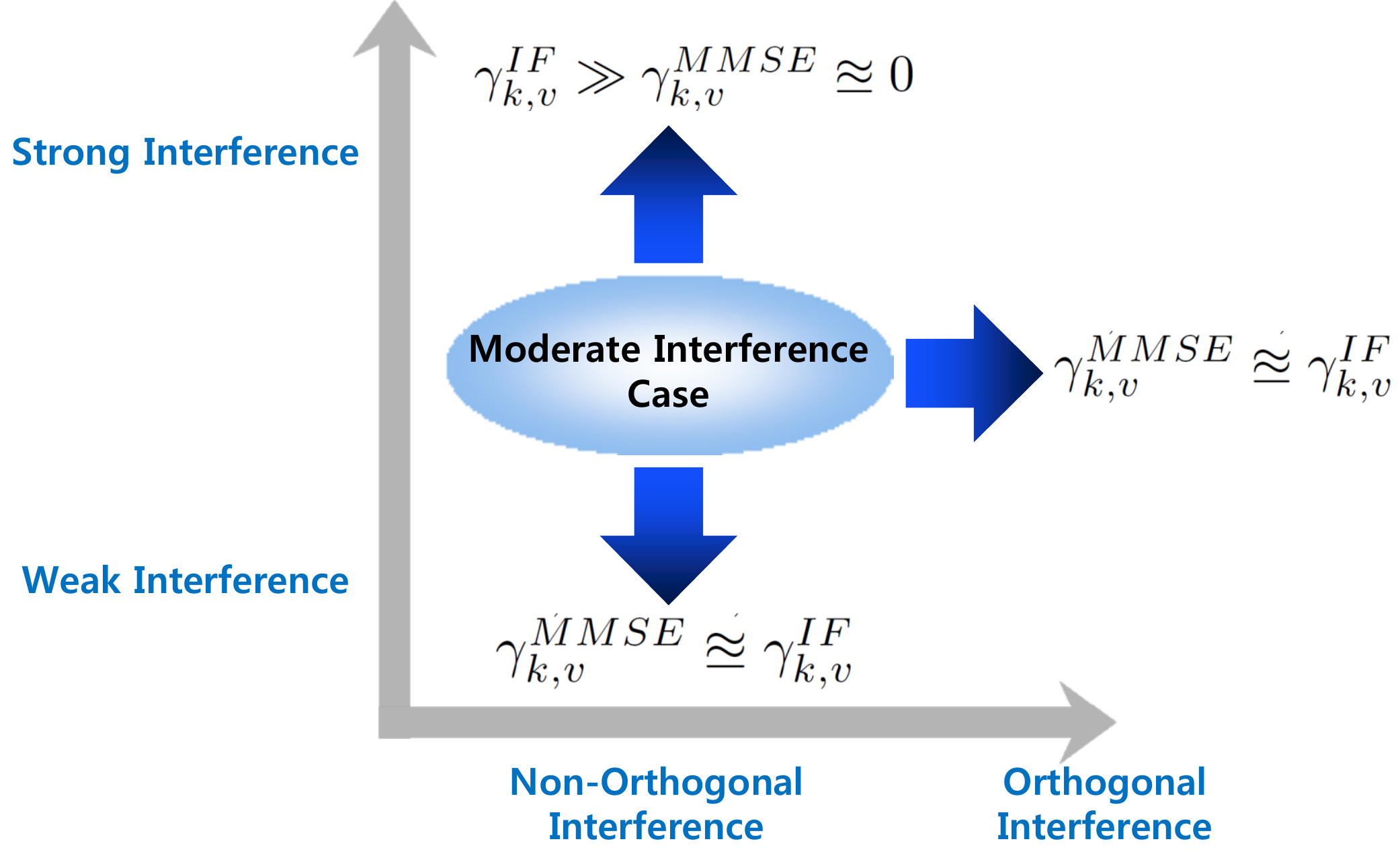}
\caption{Three Extreme Cases: Weak Interference, Strong Interference, Orthogonal Interference, and the relation between the lower and upper bounds} \label{fig:1}
\end{center}
\end{figure}
In this subsection, we show that the constant $\beta$-based approach shown in (\ref{eq:9})  has an inherent drawback in the case of strong interference. To this end, we take an information-theoretic view of two bounds $\gamma^{MMSE}_{k,v}$ and $\gamma^{IF}_{k,v}$ in the three extreme interference cases: weak interference, strong interference and orthogonal interference. The interference environments can be split into two regimes: moderate and extreme interference regimes, as shown in Fig. \ref{fig:1}. We note that the lower bound changes as the interference condition moves from the moderate case to the three extreme cases while the upper bound remains unchanged. In order to simplify analysis, we assume a single-layer transmission  for each user, i.e., $V_i=1$ though results can be generalized to the case with $V_i\ge1$. In this case, the system model in (\ref{eq:1}) reduces to
\bea \label{eq:10} \rv_k
= \hv^1_k x^1_k +\hv^2_k x^2_k + \nv_k.
\eea 

Let us first consider the orthogonal interference. When the subspace spanned by the serving channel $\hv^1_k$ becomes more orthogonal to that of the interfering channel $\hv^2_k$, the MMSE receiver performs asymptotically the same as the genie-aided IF receiver. This is consistent with the fact that the lower bound $\gamma^{MMSE}_{k,v}$ in (\ref{eq:5}) is coincident with the upper bound $\gamma^{IF}_{k,v}$ in (\ref{eq:7}) when $\left(\hv^1_k\right)^{\dagger}\hv^2_k=0$.

Next suppose that the UE is physically closer to the serving BS than to the interfering BS and, hence, the received signal from the interfering BS is much smaller than that from the serving BS. 
Under this weak interference condition, treating interference as noise is near-optimal in terms of system throughput \cite{Gamal:11}. Again, this is consistent with the fact that the lower bound $\gamma^{MMSE}_{k,v}$ practically coincides with the upper bound $\gamma^{IF}_{k,v}$ as the interfering part $\hv^2_k x^2_k$ in (\ref{eq:10}) becomes asymptotically negligible.  

One important feature to notice about the equation in (\ref{eq:9}) is that as long as the lower bound  $\gamma^{MMSE}_{k,v}$ is the same as the upper bound $\gamma^{IF}_{k,v}$,  (\ref{eq:9}) yields $\gamma^{ML}_{k,v}$ the same as $\gamma^{MMSE}_{k,v}$ and $\gamma^{IF}_{k,v}$ regardless of $\beta$. 
This implies that the layer separation method based on (\ref{eq:9}) is able to obtain an accurate estimate of the post-MLD SINR $\gamma^{ML}_{k,v}$ in the case of weak and orthogonal interference where  $\gamma^{ML}_{k,v}$ is the same as $\gamma^{MMSE}_{k,v}$ and $\gamma^{IF}_{k,v}$ for any value of $\beta$. 

Finally, consider the interference channels under very strong interference. The seminal work of Carleial \cite{Carleial:75} showed that very strong interference is equivalent to no interference and Sato extended the work to interference channels with strong interference \cite{Sato:81}. In other words, the message $x_k^1$ can be recovered reliably, under the strong interference, at the same rate that is achievable in the absence of interference $\hv^2_k x^2_k$. This implies that for the conventional static approach to work properly, the lower bound $\gamma^{MMSE}_{k,v}$ should converges to the upper bound $\gamma^{IF}_{k,v}$ in strong interference region. However, the equation (\ref{eq:6}) indicates that the actual lower bound $\gamma^{MMSE}_{k,v}$ decreases to zero asymptotically when the interfering part $\hv^2_k x^2_k$ becomes stronger. 
Therefore, the conventional approach based on the fixed ratio $\beta$ can lead to wrong layer separation in the strong interference case.

In the following section, we will propose a new adaptive approach, where the combining ratio is adaptively computed based on the instantaneous ISR, accounting for this misleading lower bound.
\section{ Proposed ISR-Adaptive approach to the layer separation } \label{sec:4}
In this section, we present a new adaptive approach to the layer separation for MIMO systems which overcomes the drawback of the conventional static approach. As stated above, we focus on the MIB metric.    

\subsection{Proposed Adaptive Approach}
The lower and upper bounds to the post-MLD MIB denoted by $\cM^{ML}_k$ are given by mapping the corresponding SINR bound to an MIB value, respectively, as
\bea  \label{eq:11}
\cM^{low}_{k}=\cI_{M_c^{1}}\left(\gamma^{MMSE}_{k,v}\right),
\eea
and
\bea  \label{eq:12}
\cM^{up}_{k}=\cI_{M_c^{1}}\left(\gamma^{IF}_{k,v}\right).
\eea
where for notation brevity, we drop off the layer index $v$, i.e., $\cM^{low}_{k}=\cM^{low}_{k,v}$ and
$\cM^{up}_{k}=\cM^{up}_{k,v}$.

In this work, we use the closed-form expressions suggested in \cite{Roshni:08} (See Table 28 therein) to approximate the MIB mapping functions $\cI_{M_c}\left(\gamma\right)$. 

By applying the MIB metric $\cF(\gamma)=\cI_{M^1_c}\left(\gamma\right)$, we can rewrite the equation (\ref{eq:9}) as
\bea  \label{eq:13}
\cM^{ML}_k&=& (1- \beta_{ISR})\cM^{low}_{k} + \beta_{ISR}\cM^{up}_{k},
\eea
where we use the subscript ${ISR}$ in order to emphasize the dependency on the ISR. 
As seen from the definition, the parameter $\beta_{ISR}$ can not be larger than one $\left(\beta_{ISR}\le1\right)$.

\begin{figure}
\begin{center}
\includegraphics[width=5.5in]{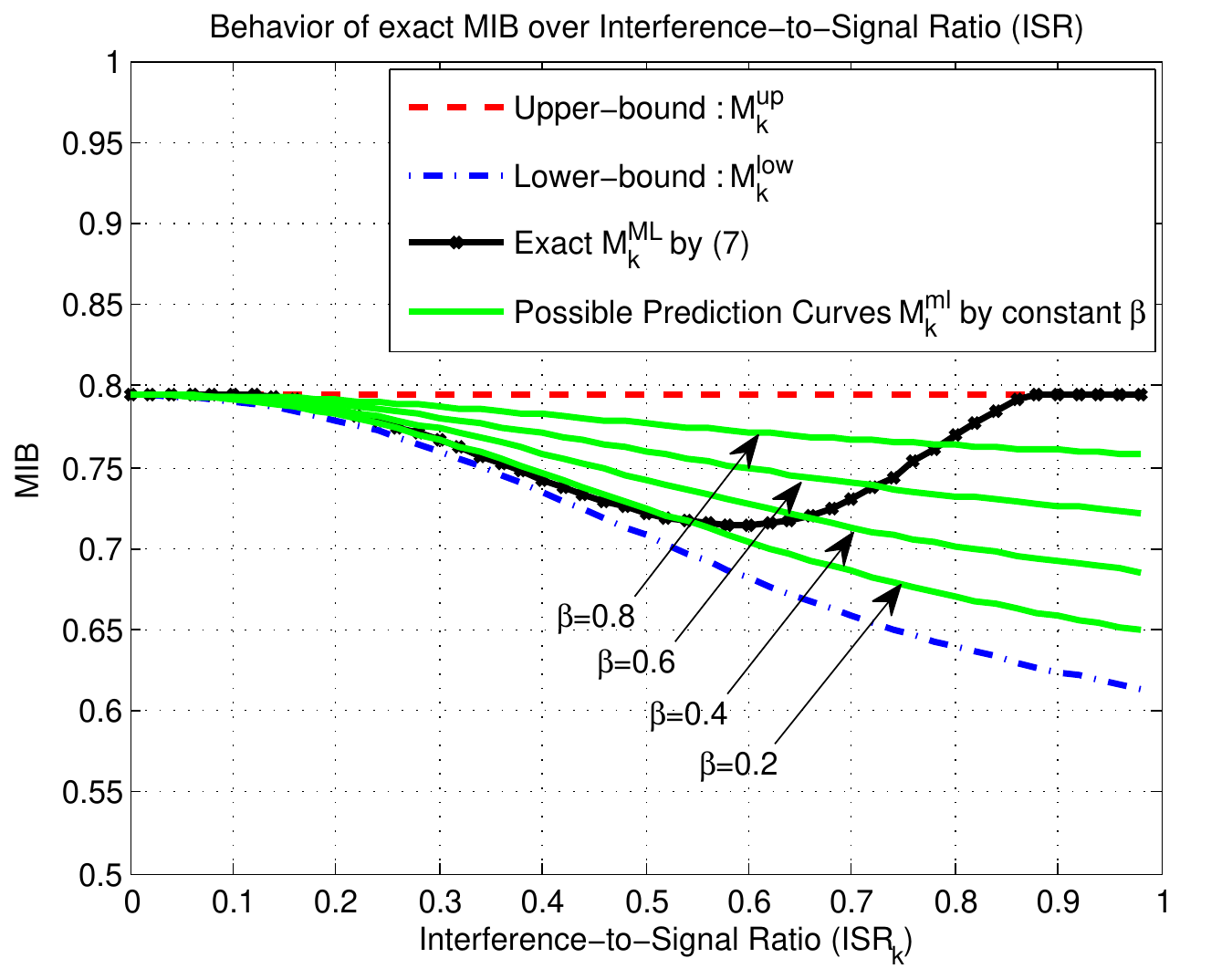}
\caption{Behavior of post-MLD MIB over Interference-to-Signal Ratio (ISR)} \label{fig:2}
\end{center}
\end{figure}
As shown in the subsection \ref{sec:3B}, the conventional static combining approach is not able to accommodate the probabilistic behavior of optimal $\beta_{ISR}$ in the strong interference scenarios. 
Fig. \ref{fig:2} depicts the exact MIB $\cM^{ML}_{k}$ given by (\ref{eq:4}) for one specific channel realization $\hv^1_k$ and $\hv^2_k$ with respect to the ISR $ISR_k$, comparing with the upper and lower bounds. Here, motivated by the Chernoff bound expression on pairwise error probability in Rayleigh fading channels, we define $ISR_k$ as 
\bea  \label{eq:15}
ISR_k=1-\exp{\left(-\frac{||\hv^2_k||}{||\hv^1_k||}\right)}.
\eea
By the definition of $ISR_k$ in (\ref{eq:15}), $ISR_k$ ranges between $0$ and $1$, i.e., $0\le ISR_k\le1$. The ratio $ISR_k$ asymptotically decreases to zero and increases to one, respectively, under the weak and the strong interference conditions.

Fig. \ref{fig:2} shows that while  the exact $\cM^{ML}_{k}$ may be well approximated by a constant $\beta$ value for low ISRs (e.g., $\beta=0.2$ for $ISR_k\le 0.5$), the optimal combining ratio $\beta$ continues to increase to one with increasing $ISR_k$. From this observation, we can conclude that the conventional static approach based on the fixed $\beta$ can not capture the behavior of the exact MIB $\cM^{ML}_{k}$ especially in high ISR regimes. 
The observation suggests that the combining parameter $\beta_{ISR}$ needs to be adaptively chosen according to the instantaneous ISR. 

To this end, we propose the following ISR-adaptive parameterization of $\beta_{ISR}$
\bea  \label{eq:14}
\beta_{ISR}=\cB\left(ISR_k,MCS_1,M^2_c\right),
\eea
where $M^2_c$ denotes the modulation order of interfering BS. 

\begin{figure}
\begin{center}
\includegraphics[width=5.5in]{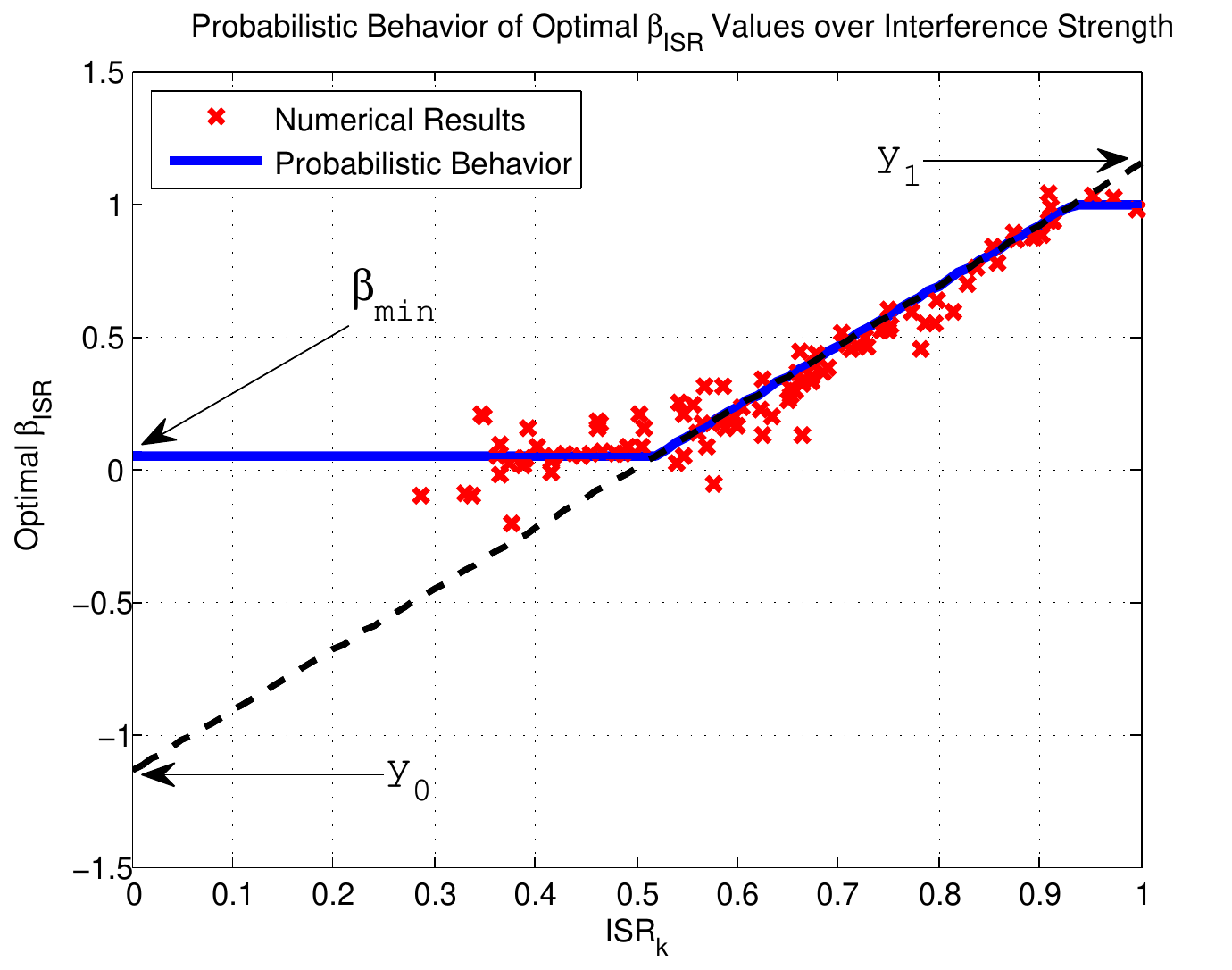}
\caption{Probabilistic Behavior of Optimal $\beta_{ISR}$ over the ISR for $MCS_1=9$ and $M^2_c=4$QAM } \label{fig:3}
\end{center}
\end{figure}
In order to characterize the probabilistic behavior of the optimal $\beta_{ISR}$ in Equation (\ref{eq:13}) over  $ISR_k$, we present the numerical results for the case of $MCS_1=9$ and $M^2_c=4$QAM  in Fig. \ref{fig:3}. 
By applying Monte-Carlo simulations to Equation (\ref{eq:3}), we first obtain the exact MIB $\cM^{ML}_k$ from Equation (\ref{eq:4}) for given channel realization. We also compute the corresponding two bounds $\cM^{low}_{k}$ and $\cM^{up}_{k}$ from (\ref{eq:5}), (\ref{eq:7}), (\ref{eq:11}), and (\ref{eq:12}). Then, we can find the optimal $\beta_{ISR}$ value which satisfies (\ref{eq:13}). Fig. \ref{fig:3} depicts the optimal $\beta_{ISR}$ values with respect to $ISR_k$ for $100$ channel realizations of $\hv^1_k$ and $\hv^2_k$ which achieve the MIB values corresponding to the target BLER in the AWGN reference curve of the involved $MCS_1$. In this figure, the target BLER of interest is $10^{-1.1}$ to $10^{-0.9}$ and accordingly, the related SINR range is $-2$ dB to $10$ dB. 

In Fig. \ref{fig:3}, we can see that the optimal $\beta_{ISR}$ is well approximated by a single constant value for low ISRs and increases linearly to one as $ISR_k$ grows. This probabilistic behavior of the optimal $\beta_{ISR}$ is in line with the discussion made above with  Fig. \ref{fig:2}. Then, it follows from the observation that the optimal $\beta_{ISR}$ behavior can be approximated by the following piecewise approximation, represented by the solid blue line in Fig. \ref{fig:3},
\bea  \label{eq:16}
\beta_{ISR}=\max \left\{ \min \left\{\left(y_{1}-y_{0}\right)ISR_k+y_{0}, 1\right\}, \beta_{min} \right\},
\eea
where as depicted in the figure, link abstraction model parameters $y_0$, $y_1$ and $\beta_{min}$ are  for optimization.

We notice that the simplification of $\beta_{ISR}$ in (\ref{eq:16}) can be also justified by the fact that as shown in the subsection \ref{sec:3B}, the accuracy of layer separation based on (\ref{eq:9}) becomes less sensitive with decreasing ISR since the lower bound $\gamma^{MMSE}_{k,v}$ approaches the upper bound $\gamma^{IF}_{k,v}$ in the low ISR region.  
\subsection{Generalization to the case of $V_1\ge 1$ and $V_2\ge 1$}
We can extend the proposed method for arbitrary numbers of transmission layers $V_1\ge 1$ and $V_2\ge 1$. In this subsection, as an example we provide a brief description of generalization to the case of $V_1\le 2$ and $V_2\le 2$. The layer separation of multiple-layer cases proceeds in the same fashion as the single-layer case  while the following modifications are required. The signal model can be written in the form
\bea \label{eeq:1}
\rv_k\!\!\!\! &=& \!\!\!\!\Hv^1_k  \xv^1_k +\Hv^2_k \xv^2_k + \nv_k,\\
    \!\! \!\! &=&\!\!\!\!\! \left[\hv^1_{k,1} \hv^1_{k,2} \right]\!\left[\begin{array}{c}x^{1,1}_k\\ x^{1,2}_k \end{array}\right] \! +\!\left[\hv^2_{k,1} \hv^2_{k,2} \right]\! \left[\begin{array}{c}x^{2,1}_k\\ x^{2,2}_k \end{array}\right] \!+\!\! \nv_k. 
\eea 
By focusing our attention on the first transmission layer of the desired UE, we can extend the definition of $ISR_k$ in (\ref{eq:15}) as follows: 
\bea  \label{eeq:2}
ISR_k=1-\exp{\left(-\frac{||\Hv^{2,eff}_k||_F}{||\hv^1_{k,1}||}\right)},
\eea
where $||\cdot||_F$ denotes Frobenius norm and $\Hv^{2,eff}_k$ is the effective interference channel matrix defined as $\Hv^{2,eff}_k=\left[\Hv^2_k\right]$ and $\Hv^{2,eff}_k=\left[\hv^1_{k,2},~ \Hv^2_k\right]$, respectively, for $V_1=1$ and $V_1=2$. 

\subsection{Effective SINR mapping (ESM)} 
As one codeword in a coded OFDM system is transmitted over the subcarriers which have different channel gains, we require ESM to map the post-processing MIB values across the subcarriers into a single SINR value, which is then used to estimate instantaneous BLER of the link by looking up the AWGN reference curve. 

Link performance abstraction is given as a function of MIB values  $\cM^{ML}_k$  across the subcarriers belonging to one codeword.  Once the (per-layer) post-MLD MIB values $\{\cM^{ML}_k\}_{k=1}^K$ over the $K$ sub-carriers are obtained, we compute a  mean MIB (MMIB), denoted by $\cM^{ML}_{mmib}$, as
\bea  \label{eq:17}
\cM^{ML}_{mmib}=\frac{1}{K}\sum_{k=1}^{K}\cM^{ML}_k.
\eea

Then, $\cM^{ML}_{mmib}$ can be inversely mapped to get the effective SINR
\bea   \label{eq:18}
SINR_{eff}=\cI^{-1}_{M_c^{1}}\left(\cM^{ML}_{mmib}\right).
\eea

Finally, the estimate of BLER can be obtained by mapping $SINR_{eff}$ to BLER via looking up the AWGN look-up table 
\bea  \label{eq:19}
BLER_{est}=LUT_{AWGN}\left(SINR_{eff},MCS_1\right),
\eea
where $LUT_{AWGN}\left(SNR,MCS\right)$ is the mapping function which is specific to the involved MCS and code length. The mapping functions need to be acquired in advance from LLS over AWGN channel for the all specific conditions of interest. 

It is worthy of noting that we can also use a direct MMIB to BLER relationship in order to directly map $\cM^{ML}_{mmib}$ to the estimated BLER as follows \cite{Roshni:08}
\bea  \label{eq:20}
BLER_{est}=LUT^{mmib}_{AWGN}\left(\cM^{ML}_{mmib},MCS_1\right),
\eea
where $LUT^{mmib}_{AWGN}\left(MIB,MCS\right)$ can be derived from the two functions 
$\cI_{M_c}\left(\gamma\right)$ and $LUT_{AWGN}\left(SNR,MCS\right)$

In summary, as the noise effect of SNR and INR is captured in (\ref{eq:5}) and (\ref{eq:7}), our link abstraction method needs only the table of three parameters $y_0$, $y_1$ and $\beta_{min}$ for each set\footnote{Note that MIB is dependent on the modulation level, namely, $M^1_c$ and $M^2_c$, but not the code rate. However, as will be clear in subsequent sections, the code rate, i.e., $MCS_1$, will affect the best parameters of $y_0$, $y_1$ and $\beta_{ISR}$ which will be found by training.} of $MCS_1$ and $M^2_c$ along with the MIB and AWGN reference tables, respectively, for $\cI_{M_c}\left(\gamma\right)$ and $LUT_{AWGN}\left(SNR,MCS\right)$.    
  
\section{Discussion and Numerical Results} \label{sec:5}
In this section, we evaluate the performance of the proposed ISR-adaptive link abstraction method, comparing with the conventional static method to demonstrate the efficacy of our proposed method. In addition, the proposed link abstraction method is applied to the SLS. To this end, we need to tune the model parameters $y_0$, $y_1$ and $\beta_{min}$ by training the proposed link abstraction model in (\ref{eq:16}) under the advanced IAC receiver with the closed-loop 2-by-2 MIMO configuration specified for LTE systems. Unless otherwise stated, we assume LTE 3GPP specifications (series 36) as the baseline for our following discussion and simulations  \cite{211:12} \cite{212:12} \cite{213:12}.  

The aim of link model training is two-folded. On one hand,  the training is considered as a process of tuning the model parameters to capture non-ideal effects in the link performance abstraction, including non-linear MMIB procedure in (\ref{eq:17}) and  (\ref{eq:18}) and non-Gaussian interference against the use of AWGN reference curves in (\ref{eq:19}). On the other hand, the training aims to avoid overestimation of link performance, taking into account practical implementation issues in the IAC receiver. For example, in order to reduce the receiver complexity, we assume that the max-log approximation is applied both for demodulation and decoding \cite{Hochwald:03} so that we can avoid the logarithm of a sum of exponential functions in computation of LLR. 
In this case, the link abstraction based on the theoretical MIB will overestimate link performance and thus we need to tune the parameter $\beta_{ISR}$ obtained by the MIB analysis in Section \ref{sec:4}.

\begin{algorithm}
\caption{Link Model Training$\left(y^{tuned}_0,y^{tuned}_1,\beta^{tuned}_{min}\right)$}
\begin{algorithmic}[1]  
\Procedure {Fitting}{$\cS_{channel},\! \cS_{snr},\! \cS_{y_0,y_1,\beta_{min}},\! MCS_1,\! M_c^2$}
\State $min \leftarrow \infty$
\ForAll {$(y_0,y_1,\beta_{min}) \in \cS_{y_0,y_1,\beta_{min}}$}
\State $mse \leftarrow 0$
\ForAll {$\!\{\hv^1_k,\hv^2_k\}_{k=1}^{K} \!\in \!\cS_{channel}\!$ and $\sigma^2_n \!\in \! \cS_{snr}$}
\For {$k \leftarrow 1, K$}
\State $\cM^{low}_{k}\leftarrow \cI_{M_c^{1}}\left(\gamma^{MMSE}_{k,1}\leftarrow\frac{1}{\sigma^2_{k,1}}-1\right)$
\State $\cM^{up}_{k}\leftarrow \cI_{M_c^{1}}\left(\gamma^{IF}_{k,1}\leftarrow\frac{||\hv^1_k||^2}{\sigma^2_n}\right)$
\State $ISR_k \leftarrow 1-\exp{\left(-\frac{||\hv^2_k||}{||\hv^1_k||}\right)}$
\State $\beta_{ISR} \leftarrow \min \left\{\left(y_{1}-y_{0}\right)ISR_k+y_{0}, 1\right\}$
\State $\beta_{ISR} \leftarrow \max \left\{\beta_{ISR}, \beta_{min}\right\}$
\State $\cM^{ML}_k\! \leftarrow(1-\beta_{ISR})\!\cM^{low}_{k}\!\!+\!\!\beta_{ISR}\cM^{up}_{k}$
\EndFor
\State $\cM^{ML}_{mmib}\leftarrow\frac{1}{K}\sum_{k=1}^{K}\cM^{ML}_k$
\State $SINR_{eff}\leftarrow \cI^{-1}_{M_c^{1}}\left(\cM^{ML}_{mmib}\right)$
\State $BLER_{est}\!\!\leftarrow \!\!LUT_{AWGN}\!\!\left(SINR_{eff}\!,\!MCS_1\!\right)\!\!$
\State $mse\!\! \leftarrow \!\!  mse+\! |\!\log{BLER_{est}}\!\!-\!\log{BLER_{monte}}|^2\!$
\EndFor
\If {$min >mse$}
\State $min \leftarrow mse$
\State $\left(y^{tuned}_0,y^{tuned}_1,\beta^{tuned}_{min}\right) \leftarrow \left(y_0,y_1,\beta_{min}\right)$
\EndIf
\EndFor
\State \textbf{return} $\left(y^{tuned}_0,y^{tuned}_1,\beta^{tuned}_{min}\right)$
\EndProcedure
\end{algorithmic}  \label{al:1}
\end{algorithm}
As a result, the training allows the link abstraction model to have the best model parameters for minimizing the error between the estimated BLER $BLER_{est}$ given by Equation (\ref{eq:19}) and the actual BLER obtained from link-level simulations. The tuned parameter $\beta_{ISR}^{tuned}$ will be obtained through numerical fitting. The details are as outlined by the pseudo-code shown in Algorithm \ref{al:1}. Here, the simulated BLER, denoted by $BLER_{monte}$, is obtained  by Monte-Carlo simulation from LLS over the SNR region of interest $\cS_{snr}$ for each realization $\{\hv^1_k,\hv^2_k\}_{k=1}^{K}$ of 100 randomly generated OFDM channel realizations $\cS_{channel}$. 
In order to avoid exhaustive search over the entire search space $\cS_{y_0,y_1,\beta_{min}}$, we apply an iterative 2D-directed search method followed by 3D-directed search. Each step of 2D search consists of searching nine locations, around the point $(y_0,y_1)$ in the $y_0$-$y_1$ plane,  given by pairs of $\{y_0,y_0\pm\Delta\}$ and $\{y_1,y_1\pm\Delta\}$.  We start with the origin $(y_0,y_1)=(0,0)$ and $\Delta=1$  with setting $\beta^{tuned}_{min}=-\infty$ to find the location with the minimum MSE and make it the new origin, denoted as $(\hat{y}^{tuned}_0,\hat{y}_1^{tuned})$. This procedure continues for the current step size $\Delta$ until the new origin is the same as the previous origin. We then repeat search  with the new step size $\Delta=\Delta/10$  until $\Delta=0.01$. 
Finally, we perform similar 3D search with the origin $\left(y_0,y_1, \beta_{min}\right)=\left(\hat{y}^{tuned}_0,\hat{y}_1^{tuned},0\right)$ but with the fixed step size $\Delta=0.01$, resulting in the best parameters $y^{tuned}_0$, $y^{tuned}_1$, and $\beta^{tuned}_{min}$..

\begin{figure}
\begin{center}
\includegraphics[width=5.5in]{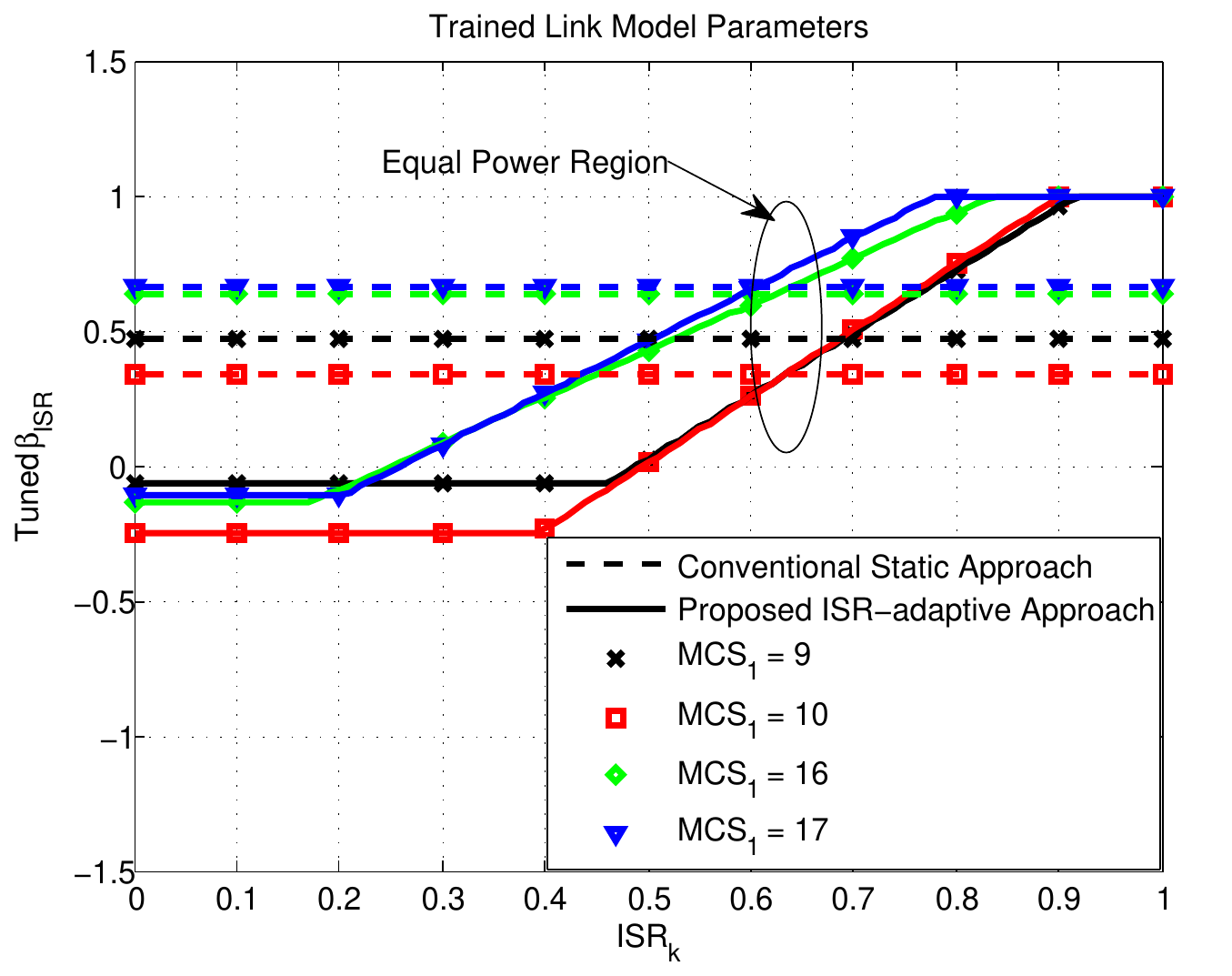}
\caption{Comparison of the trained link model parameters between Adaptive and Static Approach for $M_c^2=4QAM$} \label{fig:4}
\end{center}
\end{figure}
An example illustration of the trained link model $\beta^{tuned}_{ISR}$ with the tuned parameters  $\left(y^{tuned}_0,y^{tuned}_1,\beta^{tuned}_{min}\right)$ is given in Fig. \ref{fig:4} for a few representative MCS$_1$'s with $M_c^2=4QAM$.  For comparison purpose, we also add the plots corresponding to the conventional static approach. As shown in the figure, the two lines of adaptive and fixed model parameters intersect each other in an equal power region around $ISR_k=1-\exp(-1)=0.632$. This observation tells us that the conventional static approach tends to find the parameter $\beta$ optimal for the case of $||\hv^1_k||\approx||\hv^2_k||$.   

Let us now compare the tuned $\beta^{tuned}_{ISR}$  in Fig. \ref{fig:4}, corresponding to  the case of MCS$_1=9$, with the probabilistic behavior of optimal $\beta_{ISR}$ values given in Fig. \ref{fig:3}. Recall that they are derived independently.  From the comparison, we can see  that the tuned $\beta^{tuned}_{ISR}$ matches well with  the probabilistic behavior of optimal $\beta_{ISR}$ values, justifying our analysis in Section \ref{sec:4}.  What we should emphasize here is that the training can cause the tuned $\beta^{tuned}_{ISR}$ to be slightly lower than the optimal $\beta_{ISR}$ due to the non-ideal implementations of detection and decoding. Fig. \ref{fig:4} shows that the resultant tuned parameters $\beta^{tuned}_{ISR}$ can become less than zero at low ISRs. It is worthwhile to note that from the equation (\ref{eq:13}), a negative $\beta^{tuned}_{ISR}$ does not mean that the value of $\cM^{ML}_k$ is negative, but means that $\cM^{ML}_k$ is smaller than $\cM^{low}_{k}$.

The optimal parameters $\left(y^{tuned}_0,y^{tuned}_1,\beta_{ISR}^{tuned}\right)$ would guarantee that with increasing number of channel realizations and noise realizations for the Monte-Carlo simulations, the simulated BLER converges asymptotically to the predicted BLER, i.e.,
\bea \label{eq:21}
BLER_{monte}\approx BLER_{est}.
\eea   

\begin{algorithm}
\caption{Link Performance Abstraction}
\begin{algorithmic}[1]  
\Procedure {Abstraction}{$\{\hv^1_k,\hv^2_k\}_{k=1}^{K}, \sigma^2_n,MCS_1, M_C^2$}
\State $\left(y^{tuned}_0,y^{tuned}_1,\beta^{tuned}_{min}\right)\leftarrow \beta^{tuned}_{ISR}\left(MCS_1, M_C^2\right)$
\For {$k \leftarrow 1, K$}
\State $\cM^{low}_{k}\leftarrow \cI_{M_c^{1}}\left(\gamma^{MMSE}_{k,1}\leftarrow\frac{1}{\sigma^2_{k,1}}-1\right)$
\State $\cM^{up}_{k}\leftarrow \cI_{M_c^{1}}\left(\gamma^{IF}_{k,1}\leftarrow\frac{||\hv^1_k||^2}{\sigma^2_n}\right)$
\State $ISR_k \leftarrow 1-\exp{\left(-\frac{||\hv^2_k||}{||\hv^1_k||}\right)}$
\State $\beta_{ISR} \leftarrow \min \left\{\left(y^{tuned}_{1}-y^{tuned}_{0}\right)ISR_k+y^{tuned}_{0}, 1\right\}$
\State $\beta_{ISR} \leftarrow \max \left\{\beta_{ISR}, \beta^{tuned}_{min}\right\}$
\State $\cM^{ML}_k\! \leftarrow(1-\beta_{ISR})\!\cM^{low}_{k}\!\!+\!\!\beta_{ISR}\cM^{up}_{k}$
\EndFor
\State $\cM^{ML}_{mmib}\leftarrow\frac{1}{K}\sum_{k=1}^{K}\cM^{ML}_k$
\State $SINR_{eff}\leftarrow \cI^{-1}_{M_c^{1}}\left(\cM^{ML}_{mmib}\right)$
\State {$BLER_{est}\leftarrow LUT_{AWGN}\left(SINR_{eff},MCS_1\right)$ }
\State \textbf{return} $BLER_{est}$
\EndProcedure
\end{algorithmic} \label{al:2}
\end{algorithm}
In what follows, we will compare the predicted $BLER_{est}$ with the simulated $BLER_{monte}$ by using new channel realizations generated independently of $\cS_{channel}$ used for tunning $\beta^{tuned}_{ISR}$ in Algorithm \ref{al:1}. The details of the PHY abstraction are given in Algorithm \ref{al:2}. 

\begin{figure}
\begin{center}
\includegraphics[width=5.5in]{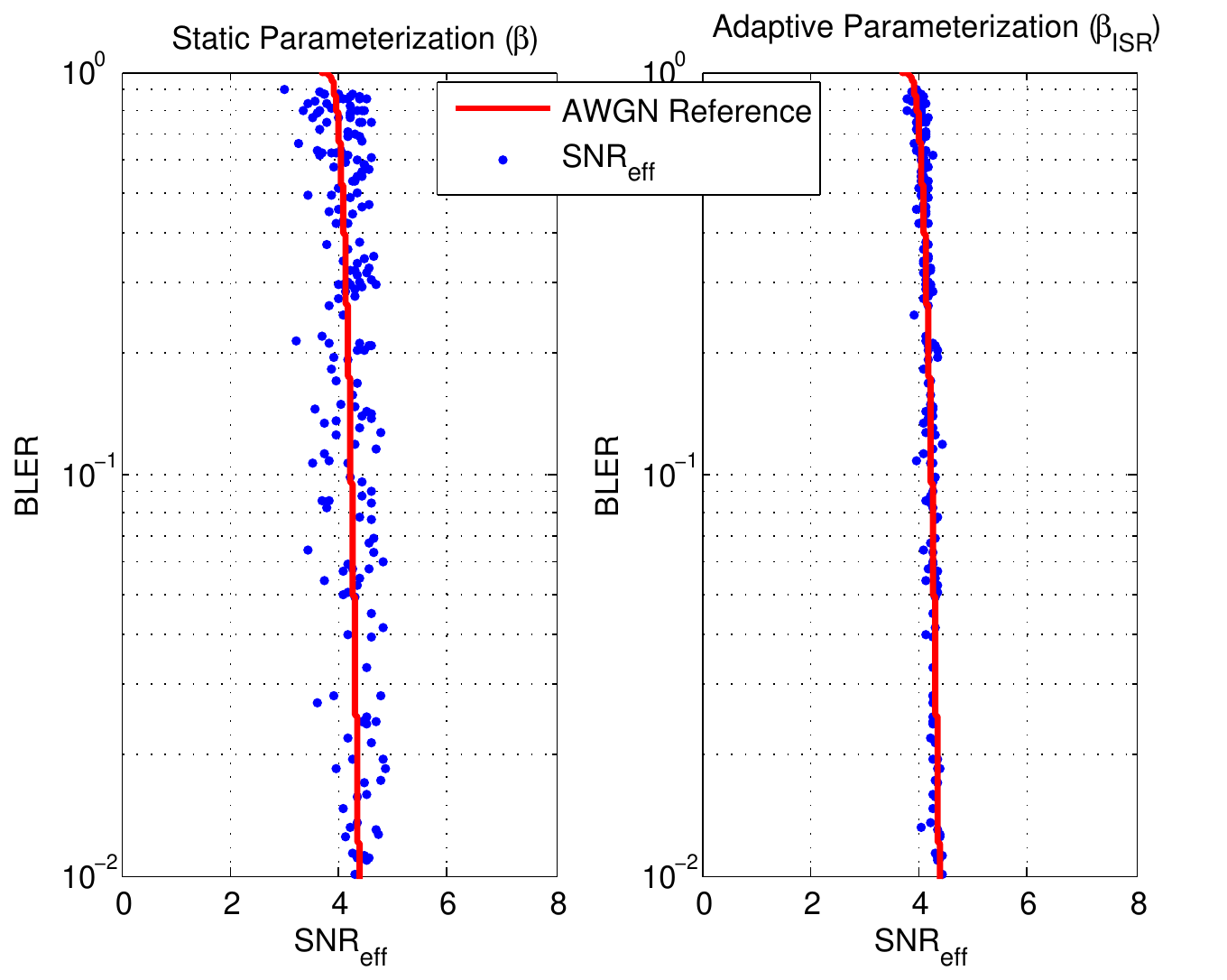}
\caption{Performance Comparison Between Static and Adaptive Approaches for  $MCS_1=9$ and $M^2_c=4QAM$ in 2x2 IAC} \label{fig:5}
\end{center}
\end{figure}
\begin{figure}
\begin{center}
\includegraphics[width=5.5in]{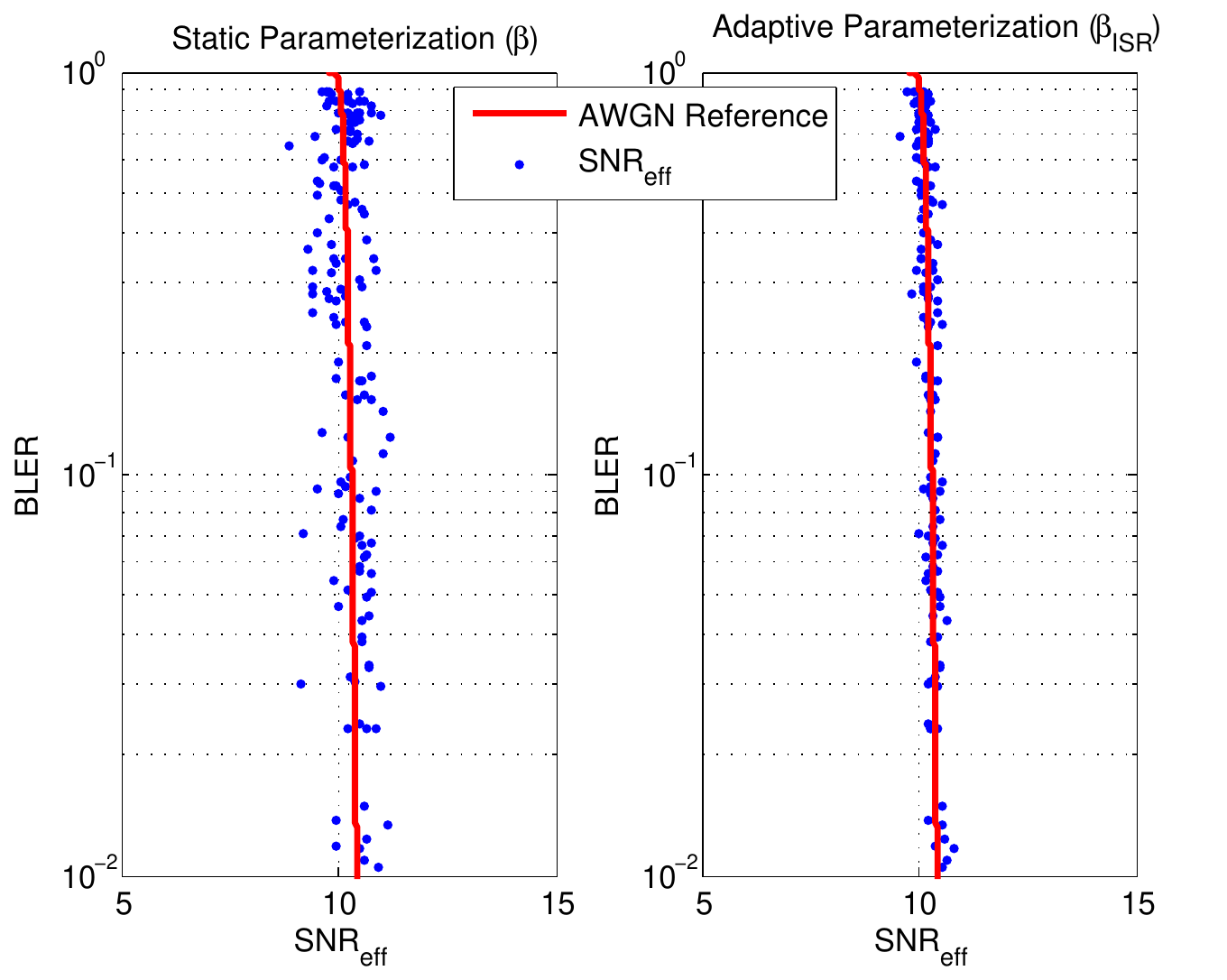}
\caption{Performance Comparison Between Static and Adaptive Approaches for $MCS_1=17$ and $M^2_c=16QAM$ in 2x2 IAC} \label{fig:6}
\end{center}
\end{figure}
Fig. \ref{fig:5} and \ref{fig:6} compare the prediction accuracy of the proposed approach with that of the conventional approach for the IAC with $V_1=1$ and $V_2=1$, denoted by  2x2 IAC, where the BSs use a combination of $MCS_1=9$ and $M^2_c=4QAM$, and $MCS_1=17$ and $M^2_c=16QAM$, respectively.  
The AWGN reference curve plots the mapping function $LUT_{AWGN}\left(SNR,MCS\right)$ corresponding to the involved $MCS_1$ while each blue dot marks the coordinate of  $\left(SINR_{eff},BLER_{monte}\right)$ for a different channel state of $\{\hv^1_k,\hv^2_k\}_{k=1}^{K} \in \cS^{new}_{channel}$ and $\sigma^2_n \in  \cS^{new}_{snr}$. Therefore, the accuracy of the abstraction can be measured by the difference between $SINR_{eff}$ and $SINR_{awgn}=LUT_{AWGN}^{-1}\left(BLER_{monte},MCS_1\right)$.
In other words, the closer the dots approach the AWGN reference curve in horizontal distance (corresponding to the distance between the two points $\left(SINR_{eff},BLER_{monte}\right)$ and $\left(SINR_{awgn},BLER_{monte}\right)$), the more accurate prediction is achieved by the link abstraction method.  The simulation results show that a substantial improvement in the prediction accuracy can be achieved if the ISR is taken into account in parameterization of combining ratio $\beta$.  Although extended simulation results for arbitrary numbers of transmission layers $V_1\ge 1$ and $V_2\ge 1$ are not presented in this paper due to the lack of space, they show that the presented approach works well for those cases.

\begin{figure}
\begin{center}
\includegraphics[width=5.5in]{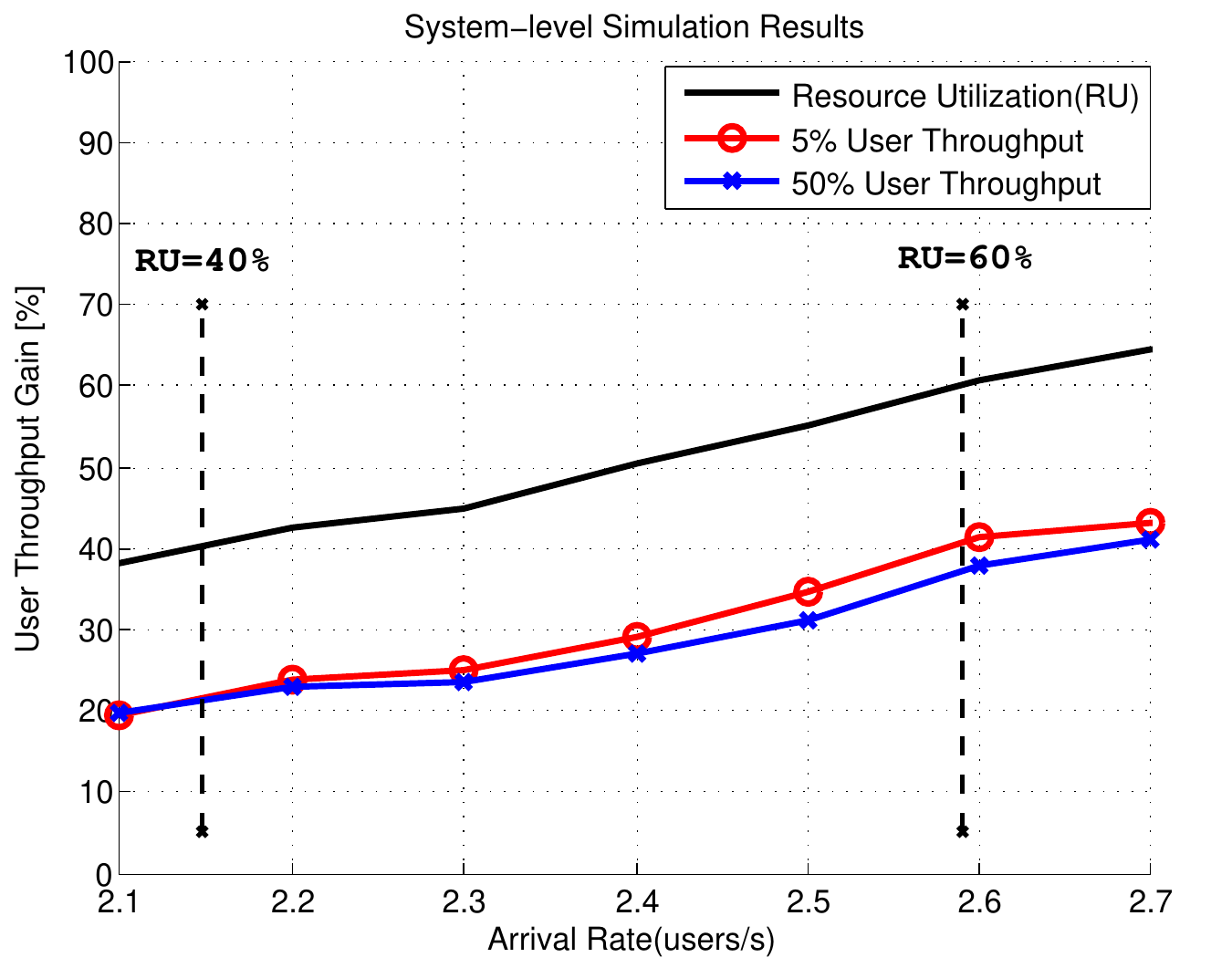}
\caption{Throughput Improvement by the advanced IAC receiver over the baseline LTE receiver} \label{fig:7}
\end{center}
\end{figure}

\begin{table}
\small
\begin{center}
\caption{Evaluation Assumptions for the system-level simulations} \label{table:1}
\begin{tabular}{|c|c|}
\hline
Parameter & Assumption \\
\hline
\hline
Cellular Layout	 & Hexagonal Grid \\
             	 & 19 cell sites, 3 sectors per site \\
\hline
Duplex	 & FDD \\
\hline
Carrier frequency &	2.0GHz\\
\hline
System  bandwidth	 & 10MHz \\
\hline
FFT size	 & 1024 \\
\hline
Subcarrier separation	 & 15KHz \\
\hline
Resource Allocation & 50RB \\
\hline
Number of UEs & 10 UEs per sector \\
\hline
Transmission mode & TM9  \\
\hline
Total BS TX power &	46dBm\\
\hline
Channel Model &	ITU M.2135 Channel Model\cite{ITU:M2135}\\
\hline
Distance-dependent path loss &	ITU Urban Macro\\
\hline
Shadowing correlation & 0.5 between cells \\
						 & 1.0 between sectors \\
\hline
Antenna pattern (Horizontal)	& $A_m=25dB$, $\varphi_{3dB}=70$ degrees\\
\hline
Antenna pattern (Vertical)	& $SLA=20dB$, $\theta_{3dB}=10$ degrees\\
\hline
Antenna Height & 	25m\\
\hline
UE antenna Height &	1.5m\\
\hline
MIMO configuration & 2Tx(0.5 lambda), Cross-polarized\\
				 & 2Rx(0.5 lambda), Cross-polarized\\
\hline
Receiver Type &	IRC Receiver(baseline LTE \cite{3GPP:829})\\
           &	IAC Receiver\\
\hline
UE noise figure	& 9dB\\
\hline
Thermal noise density	& -174dBm/Hz\\
\hline
UE speeds of interest &	3km/h\\
\hline
PCFICH &	CFI=3\\
\hline
Traffic model &	FTP traffic model 1 \cite{3GPP:814} \\
\hline
Feedback 			 &	 Feedback Period: 5 msec\\
 					 &	 Feedback Delay: 6 msec\\
\hline
Handover Margin &	 3dB\\
\hline
Scheduler &	 Proportional Fairness (PF)\\
\hline
\end{tabular}
\end{center}
\end{table}
The remaining part of this section analyzes the performance gain by the IAC receiver from system-level simulation results. The standard interference rejection combining (IRC) receiver defined in \cite{3GPP:829} has been considered as a baseline LTE  receiver. Table \ref{table:1} describes the set of simulation assumptions used for LTE system-level simulations. System throughput performance will be assessed using non-full buffer traffic model capturing bursty traffic load and time-varying interference conditions. In this work, we consider FTP traffic model 1 defined in \cite{3GPP:814} with file size of $0.5$ Mbytes. Figure \ref{fig:7} shows performance improvements by the advanced IAC receiver over the baseline LTE  receiver with respect to arrival rate. Here, resource utilization (RU) is defined as the ratio of the number of resource blocks (RBs) used by traffic to the total number of RBs available over observation time, and the user throughput is given by the file size divided by the time duration of the complete file transfer. The same traffic should be simulated for evaluating both the IAC receiver and the baseline LTE receiver. Meanwhile, the simulations are run for various arrival rates in order to cover the range of RU specified in \cite{MediaTek:13} from a low RU of $40\%$ to a high RU of $60\%$. As can be seen from Figure \ref{fig:7}, the advanced IAC receiver can achieve a throughput gain of $20\%$ at the low RU, and the gain increases with RU. Thus, the gain goes up to $40\%$ at the high RU. Note that the throughput gain of the IAC receiver comes from the joint detection of the serving and interfering signals.
\section{Conclusion} \label{sec:6}
In this paper,  we have investigated link performance abstraction of the IAC systems employing the maximum-likelihood detector in multicell interfering networks. The work was inspired by our observation that the conventional approach based on the fixed parameterization can lead to wrong link abstraction in the strong interference case. We have proposed the adaptive link abstraction model relying on the interference-to-signal ratio (ISR). The proposed ISR-adaptive strategy outperforms the conventional static approach in terms of the BLER prediction accuracy by overcoming the drawback of the conventional static strategy in high ISR region, which proves that the proposed  link abstraction model can improve the CQI calculation of the future IAC systems.
The system-level simulation results show that the advanced IAC receiver achieves significant throughput improvements over the baseline LTE  receiver in the interference-limited LTE environment.

\bibliographystyle{ieeetr}
\bibliography{AZREF}

\begin{thebibliography}{10}

\bibitem{Bea:12}
J.~H. Bae, S.~Kim, J.~Lee, and I.~Kang, ``Advanced downlink mu-mimo receiver
  for 3gpp lte-a,'' in {\em Communications (ICC), 2012 IEEE International
  Conference on}, pp.~7004--7008, june 2012.

\bibitem{Jungwon:12}
J.~Lee, H.~Kwon, and I.~Kang, ``Interference mitigation in mimo interference
  channel via successive single-user soft decoding,'' in {\em Information
  Theory and Applications Workshop (ITA), 2012}, pp.~180 --185, feb. 2012.

\bibitem{MediaTek:13}
3rd Generation Partnership Project~(3GPP), ``{Study on Network-Assisted
  Interference Cancellation and Suppression for LTE},'' {\em TSG RAN Meeting
  59}, vol.~RP-130404, Feburuary 2013.

\bibitem{Moon:12}
S.-H. Moon, K.-J. Lee, J.~Kim, and I.~Lee, ``Link performance estimation
  techniques for mimo-ofdm systems with maximum likelihood receiver,'' {\em
  Wireless Communications, IEEE Transactions on}, vol.~11, pp.~1808 --1816, may
  2012.

\bibitem{3GPP:11}
3rd Generation Partnership Project~(3GPP), ``{Coordinated multi-point operation
  for LTE physical layer aspects},'' {\em TR 36.819 v11.0.0}, Semtember 2011.

\bibitem{dahlman20114g}
E.~Dahlman, S.~Parkvall, and J.~Skold, {\em 4G: LTE/LTE-Advanced for Mobile
  Broadband}.
\newblock Elsevier Science, 2011.

\bibitem{Zehavi:92}
E.~Zehavi, ``{8-PSK trellis codes for a Rayleigh channel},'' {\em IEEE
  Transactions on Communications}, vol.~40, pp.~873--884, May 1992.

\bibitem{Caire:98}
G.~Caire, G.~Taricco, and E.~Biglieri, ``{Bit-Interleaved Coded Modulation},''
  {\em IEEE Transactions on Information Theory}, vol.~44, pp.~927--946, May
  1998.

\bibitem{Inkyu:06}
I.~Lee and C.-E. Sundberg, ``{Wireless OFDM systems with Multiple Transmit and
  Receive Antennas with Bit Interleaved Coded Modulation},'' {\em IEEE Wireless
  Communications}, pp.~80--87, June 2006.

\bibitem{Roshni:08}
R.~Srinivasan, ``{IEEE 802.16m Evaluation Methodology Document (EMD)},'' {\em
  IEEE 802.16m-08/004r2}, July 2008.

\bibitem{3GPP:829}
3rd Generation Partnership Project~(3GPP), ``{Technical Specification Group
  Radio Access Network; Enhanced performance requirement for LTE User Equipment
  (UE)},'' {\em TR 36.829 v11.1.0}, December 2012.

\bibitem{211:12}
3rd Generation Partnership Project~(3GPP), ``{Evolved Universal Terrestrial
  Radio Access (E-UTRA); Physical Channels and Modulation (Release 10)},'' {\em
  TS 36.211}, March 2012.

\bibitem{Biglieri:00a}
E.~Biglieri, G.~Taricco, and E.~Viterbo, ``{Bit-Interleaved Time-Space Codes
  for Fading Channels},'' in {\em Conference on Information Sciences and
  Systems}, pp.~WA4.1--WA4.6, March 2000.

\bibitem{Stanford:379}
J.~M. Cioffi, {\em {379A Class note: Signal Processing and detection}}.
\newblock Stanford Univ.

\bibitem{Gamal:11}
A.~E. Gamal and Y.-H. Kim, {\em {Network Information Theory}}.
\newblock Cambridge University Press, 2011.

\bibitem{Carleial:75}
A.~Carleial, ``A case where interference does not reduce capacity (corresp.),''
  {\em Information Theory, IEEE Transactions on}, vol.~21, pp.~569 -- 570, sep
  1975.

\bibitem{Sato:81}
H.~Sato, ``The capacity of the gaussian interference channel under strong
  interference (corresp.),'' {\em Information Theory, IEEE Transactions on},
  vol.~27, pp.~786 -- 788, nov 1981.

\bibitem{212:12}
3rd Generation Partnership Project~(3GPP), ``{Evolved Universal Terrestrial
  Radio Access (E-UTRA); Multiplexing and channel coding (Release 10)},'' {\em
  TS 36.212}, March 2012.

\bibitem{213:12}
3rd Generation Partnership Project~(3GPP), ``{Evolved Universal Terrestrial
  Radio Access (E-UTRA); Physical layer procedures (Release 10)},'' {\em TS
  36.213}, March 2012.

\bibitem{Hochwald:03}
B.~Hochwald and S.~T. Brink, ``{Achieving near-capacity on a multiple-antenna
  channel},'' {\em IEEE Transactions on Communications}, vol.~51, pp.~389--399,
  March 2003.

\bibitem{ITU:M2135}
R.~S. of~International Telecommunication Union (ITU-R), ``{Guidelines for
  evaluation of radio interface technologies for IMT-Advanced},'' {\em
  M.2135-1}, December 2009.

\bibitem{3GPP:814}
3rd Generation Partnership Project~(3GPP), ``{Evolved Universal Terrestrial
  Radio Access (E-UTRA); Further advancements for E-UTRA Physical Layer
  Aspects},'' {\em TR 36.814 v9.0.0}, March 2010.

\end{thebibliography}

\end{document}